\documentclass[a4paper,11pt]{article}
\usepackage{pos}
\usepackage{natbib}
\title{Diffractive single and di-hadron production at the NLO in a saturation framework}

\author*[a]{Michael Fucilla}
\author[b,c]{Andrey Grabovsky}
\author[a]{Emilie Li}
\author[d]{Lech Szymanowski}
\author[a]{Samuel Wallon}

\affiliation[a]{Université Paris-Saclay, CNRS/IN2P3, IJCLab,
   91405, Orsay, France}

\affiliation[b]{Budker Institute of Nuclear Physics, 
11, Lavrenteva avenue, 630090, Novosibirsk, Russia}

\affiliation[c]{Novosibirsk State University, 630090, 2, Pirogova street, Novosibirsk, Russia}

\affiliation[d]{National Centre for Nuclear Research (NCBJ), Pasteura 7, 02-093 Warsaw,  Poland}

\emailAdd{Michael.Fucilla@unical.it}
\emailAdd{A.V.Grabovsky@inp.nsk.su}
\emailAdd{Emilie.Li@ijclab.in2p3.fr}
\emailAdd{Lech.Szymanowski@ncbj.gov.pl}
\emailAdd{Samuel.Wallon@ijclab.in2p3.fr}

\abstract{Motivated by the need to increase the precision of theoretical predictions to test saturation physics at both the LHC and the EIC, we compute the cross-sections for the diffractive single and di-hadron production at the NLO in the shockwave formalism.}

\FullConference{The European Physical Society Conference on High Energy Physics (EPS-HEP2023)\\
 21-25 August 2023\\
Hamburg, Germany\\}

\begin{document}
\maketitle

\section{Introduction}
The study of diffractive reactions, i.e. of scattering processes characterized by the existence of a rapidity gap, is of considerable importance at modern and future accelerators, because it allows the so-called Regge-Gribov limit of QCD to be investigated. In this regime, for sufficiently high energies, the sharp increase of cross-sections with the center-of-mass energy, predicted by the BFKL approach, must be slowed down by saturation phenomena. In this work, we investigate the diffractive single and di-hadron production at small-$x$ trough the Balitsky's shockwave formalism~\cite{Balitsky:2001re} at the next-to-leading order (NLO). This new class of processes are golden channels to investigate gluon tomography in the nucleon and nuclei~\cite{Hatta:2022lzj,Marquet:2009ca, Iancu:2021rup} and widen the list of processes useful for testing saturation at the NLO~\cite{Boussarie:2014lxa,Boussarie:2016ogo,Boussarie:2019ero,Boussarie:2016bkq,Chirilli:2012jd,Beuf:2022ndu,Altinoluk:2020qet,Taels:2022tza,Roy:2019hwr,Caucal:2021ent,Caucal:2022ulg,Caucal:2023fsf,Bergabo:2022zhe,Iancu:2022gpw,Mantysaari:2022bsp,Mantysaari:2022kdm,Beuf:2022kyp,Taels:2023czt}. 

\subsection*{Shockwave formalism}

The shockwave approach is an effective field theory in which the gluonic field $A$ is separated into external background fields $b$ and  internal fields $\mathcal{A}$, through a rapidity cut-off $e^\eta p^+$, where $p^{+} \sim \sqrt{s}$ is a large longitudinal momentum in the direction of a light-cone four vector $n_1^{\mu}= \frac{1}{\sqrt{2}} (1,\Vec{0},1)$ and $\eta < 0$. The external field, after being highly boosted from the target rest frame to the probe frame, takes the form 
\begin{equation}
    b^\mu (x) = b^-(x_\perp) \delta (x^+) n_2^\mu \,,
\end{equation}
where  $n_2^{\mu}= \frac{1}{\sqrt{2}} (1,\Vec{0},-1)$. An all order resummation of the interactions of a quark (or a gluon) probe with the background field is expressed in terms of a Wilson line, that represents the shockwave and is located exactly at $x^+ =0$:
\begin{equation}
    U_{\vec{z}} = \mathcal{P} \exp \left(i g \int d z^+ b^-(z)\right)\,,
\end{equation}
where $\mathcal{P}$ is the usual path ordering operator for the $+$ direction. Depending on the process and the perturbative order, the amplitude can depend on different types of operators built from Wilson's lines, whose evolution is obtained by studying the dependence on the rapidity cut-off $e^\eta$. Our processes, at  the leading order (LO), are described by a color dipole operator, which in the fundamental representation of $SU(N_c)$ takes the form:
\begin{equation}
\left[\operatorname{Tr} \left(U_1 U_2^\dag\right)-N_c\right]\left(\vec{p_1},\vec{p}_2\right) = \int d^d \vec{z}_{1} d^d \vec{z}_{2\perp} e^{- i \vec{p}_1 \cdot \vec{z}_1} e^{- i \vec{p}_2 \cdot \vec{z}_2} \left[\operatorname{Tr} \left(U_{\vec{z}_1} U_{\vec{z}_2}^\dag\right)-N_c\right]\,.
\end{equation}

\subsection*{Diffractive single and di-hadron production at the LO}
The processes that we investigate are the diffractive single hadron production, i.e.~\cite{Fucilla:2023mkl}
\begin{equation}
\label{Eq:process}
    \gamma^{(*)}(p_\gamma) + P(p_0) \rightarrow h(p_{h}) + X + P'(p_{0'}) \; ,
\end{equation}
and the diffractive di-hadron production, i.e.~~\cite{Fucilla:2022wcg}
\begin{equation}
\label{Eq:process}
    \gamma^{(*)}(p_\gamma) + P(p_0) \rightarrow h_1(p_{h_1}) + h_2(p_{h_2}) + X + P'(p_{0'}) \; ,
\end{equation}
where $\gamma^{(*)}$ is a virtual or real photon, $h$ denotes an hadron, $P$ is a nucleon or a nucleus target and $X$ is an undetected system in the final state. A rapidity gap between the system $hX$ ($h_1 h_2 X$) and $P'$ is assumed. The transverse momenta $\vec{p}_h, \vec{p}_{h_1}$ and $\vec{p}_{h_2}$ are considered large enough to provide an hard scale\footnote{In the di-hadron case the $(\vec{p}_{h_1}-\vec{p}_{h_2})^2 \gg \vec{p}_{h_1}^{\; 2} , \vec{p}_{h_2}^{\; 2}$ conditions are assumed.}. In this way, we can remain completely general with respect to the photon virtuality, $Q^2$, and the squared momenta exchanged in the $t$-channel, $t$. The processes under investigation can be described by using an hybrid high-energy and collinear factorization in which the hadronic cross-sections are obtained through a convolution between the fragmentation functions and a small-$x$ resummed cross section.
\section{NLO impact factor and hard cross-section}
At the next to leading order the impact factor has real and virtual contributions and both of them contain a dipole part, coming from all diagrams, and a double dipole part, coming only from diagrams in which the shockwave is crossed. 
The virtual contributions to the impact factor exhibit all sorts of singularities that show up in the calculation: ultraviolet (UV), infrared (IR) and rapidity divergences. Introducing the Sudakov decomposition for an emitted parton, i.e.
\begin{equation}
\label{p-sudakov}
p_p^\mu = x_p p_{\gamma}^+ n_1^\mu + \frac{\vec{p}_p^{\,2}}{2 x_p p_{\gamma}^+} n_2^\mu + p_{p,\perp}^\mu\,,
\end{equation} 
where $p=q,\bar{q},g$ for quark, anti-quark and gluon, respectively, we can classify the rapidity and IR-divergences as
\begin{itemize}
    \item \textbf{Rapidity divergence}: $x_g$ goes to zero while the value of $p_{g, \perp}$ is arbitrary but strongly suppressed with respect to $p_{\gamma}^{+} \sim \sqrt{s}$.
    \item \textbf{Collinear divergence}: $p_{g, \perp} \rightarrow (x_g/x_q) p_{q, \perp}$ (collinear to the quark line) or $p_{g, \perp} \rightarrow (x_g/x_{\bar{q}}) p_{\bar{q}, \perp}$ (collinear to the anti-quark line) while $x_g$ is arbitrary.
    \item \textbf{Soft divergence}: All components linearly vanishing (both $x_g$ and $p_{g, \perp}$ go linearly to zero). Parameterizing the transverse momenta of the gluon as $p_{g, \perp} = x_g u_{\perp}$, with $|u_{\perp}|$ fixed in the limit $x_g$ goes to zero, we can then define the soft limit as $x_g$ goes to zero with $u_{\perp}$ generic. 
    \item \textbf{Soft and collinear divergence}: Soft as defined above, but with $u_{\perp} \rightarrow (1/x_q) p_{q, \perp}$ (soft and collinear to the quark line) or $u_{\perp} \rightarrow (1/x_{\bar{q}}) p_{\bar{q}, \perp}$ (soft and collinear to the anti-quark line).
\end{itemize}
To regularize the various singularities, we introduce a lower cut-off in the variable $x_g$, i.e. we set $|x_g| < \alpha$, and use dimensional regularization for the transverse components ($d = D-2= 2+2 \epsilon$). 

\subsection*{Rapidity divergences}
The double dipole part of the virtual contribution to the impact factor contains rapidity divergences. They are the natural consequence of the separation of the gluonic field through a rapidity divide. Indeed, in this separation, if a gluon belongs to the quantum corrections to the impact factor, it cannot have an arbitrarily small longitudinal component in the $n_1^{\mu}$-direction, while its transverse momentum remains fixed. This configuration corresponds to a situation in which the invariant mass between the gluon and the quark dipole is large\footnote{Note that if the transverse momentum also vanishes linearly (soft gluon), the invariant mass of the system is not increasing.}. Such terms have to be absorbed into the renormalized Wilson operators with
the help of the B-JIMWLK equation. We thus have to use the B-JIMWLK evolution for these operators from the cutoff $\alpha$ to the rapidity divide $e^{\eta}$, by writing
\begin{equation}
\widetilde{\mathcal{U}}_{12}^\alpha  = \widetilde{\mathcal{U}}_{12}^{e^\eta} - \int_\alpha^{e^\eta} \hspace{-0.3 cm} d \rho \frac{\partial \widetilde{\mathcal{U}}_{12}}{\partial \rho} \; .
\end{equation}
This operation has the effect of replacing $\alpha$ with $e^{\eta}$ in all contributions (LO and NLO) and also of producing an NLO term that has a divergence of the type $ \ln \alpha $, which removes the rapidity divergences of the impact factor.

\subsection*{UV divergences}
The treatment of the UV divergences of this process, in the shockwave formalism, is particularly easy. We note in fact that at the LO, the impact factor does not contain any strong coupling ($g$), since the latter is entirely contained in the Wilson lines that enter the description of the target matrix element. This means that all the one-loop UV singularities associated with coupling renormalization are not included in the impact factor. At the impact factor level, UV divergences are solely related to the dressing of the external quark lines. These quarks are however on-shell and therefore, in massless QCD, their singularity structure is of the type
\begin{equation}
    \Phi_{\text{dressing}} \propto \left( \frac{1}{\epsilon_{UV}} - \frac{1}{\epsilon_{IR}} \right) \; .
\end{equation}
Setting $\epsilon_{UV} = \epsilon_{IR}$ these diagrams vanish and, at the price of mixing UV and IR divergences, we avoid to perform an explicit renormalization. 

\section{Treatment of IR-singularities}

The structure of infrared singularities, although very similar, depends on the particular process and must be discussed at cross-section level. As an example, let's consider the production of a single hadron. The infrared divergences are entirely contained in the dipole $\times$ dipole pieces of the virtual and real corrections to the cross section. At the NLO, fragmentation can occur from the quark, the anti-quark or the gluon.
\subsection*{Quark fragmentation}
\begin{figure}
\begin{picture}(430,140)
\put(50,100){\includegraphics[scale=0.18]{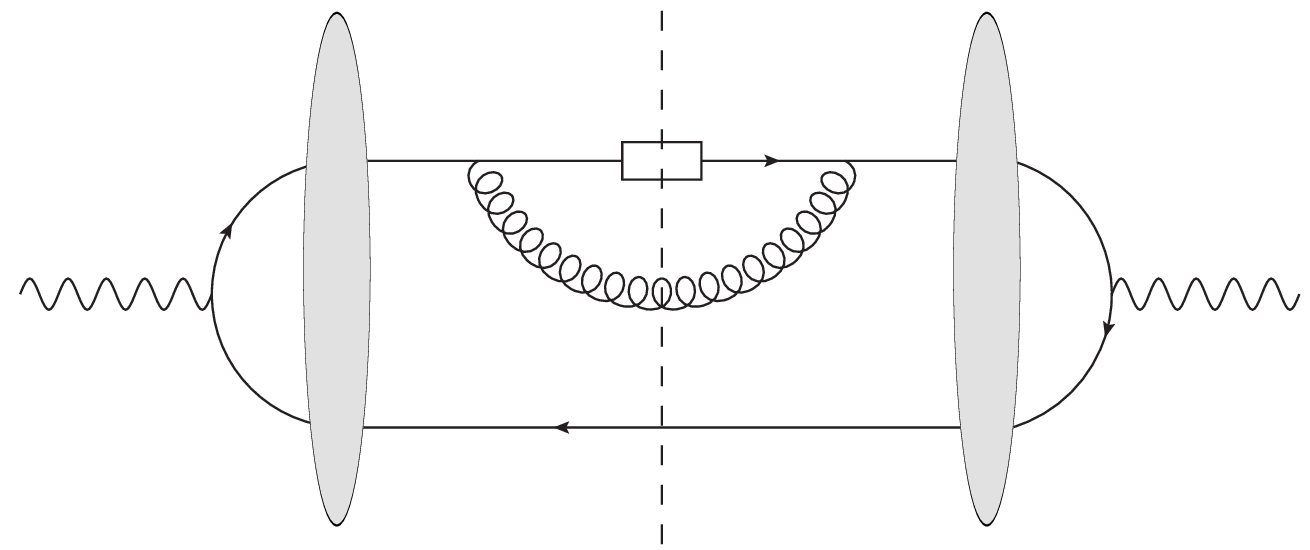}}
\put(102,80){(1)}
\put(270,100){\includegraphics[scale=0.18]{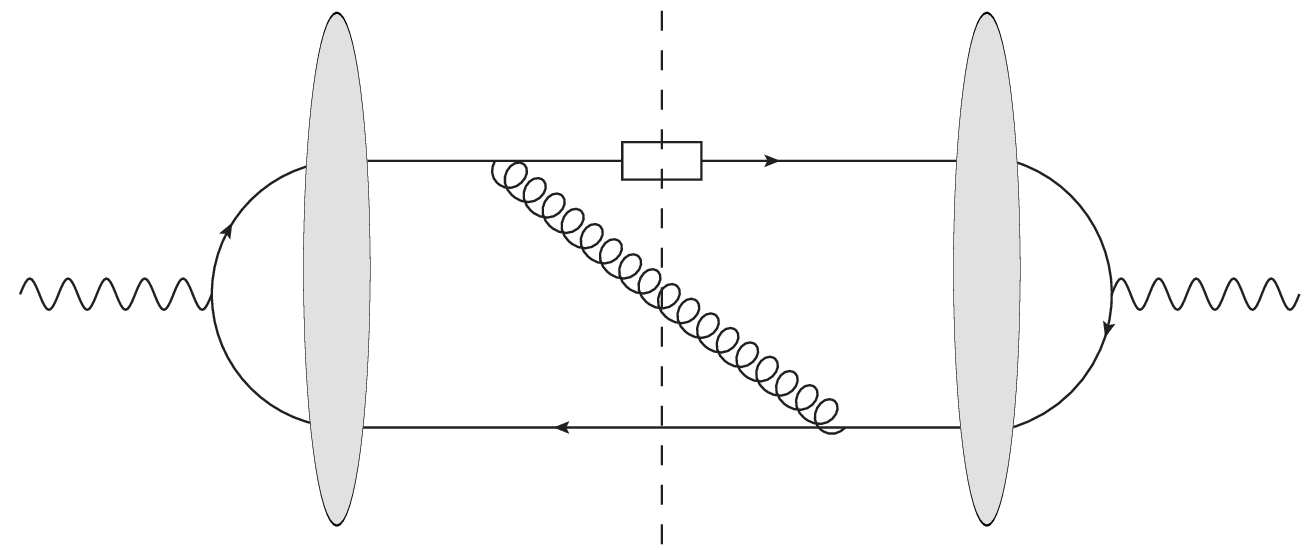}}
\put(322,80){(2)}
\put(50,20){\includegraphics[scale=0.18]{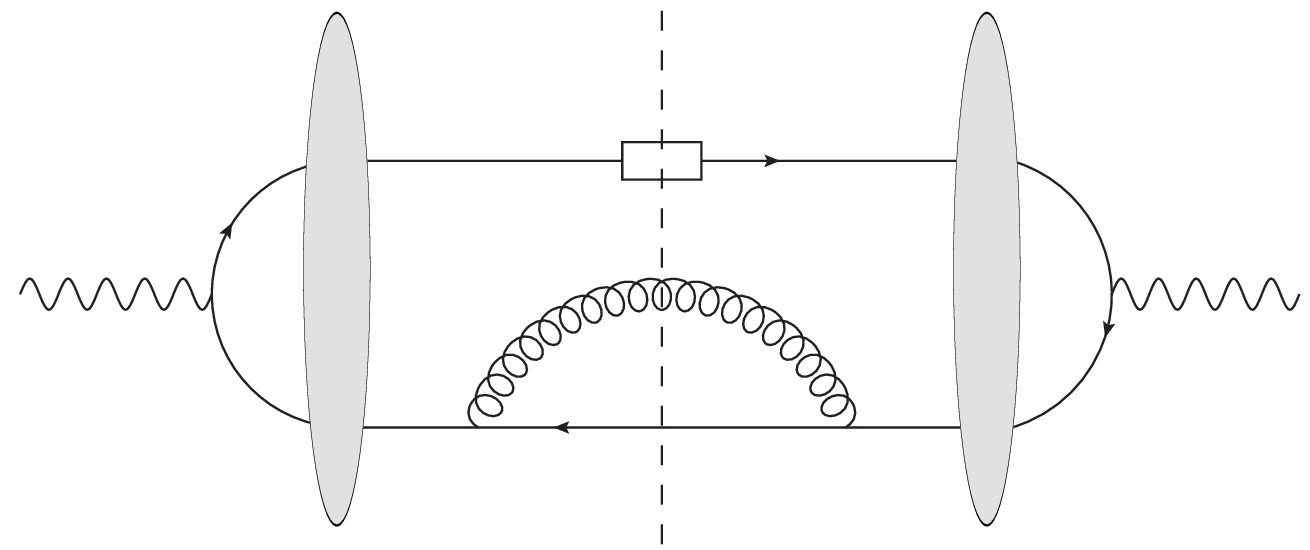}}
\put(102,5){(3)}
\put(270,20){\includegraphics[scale=0.18]{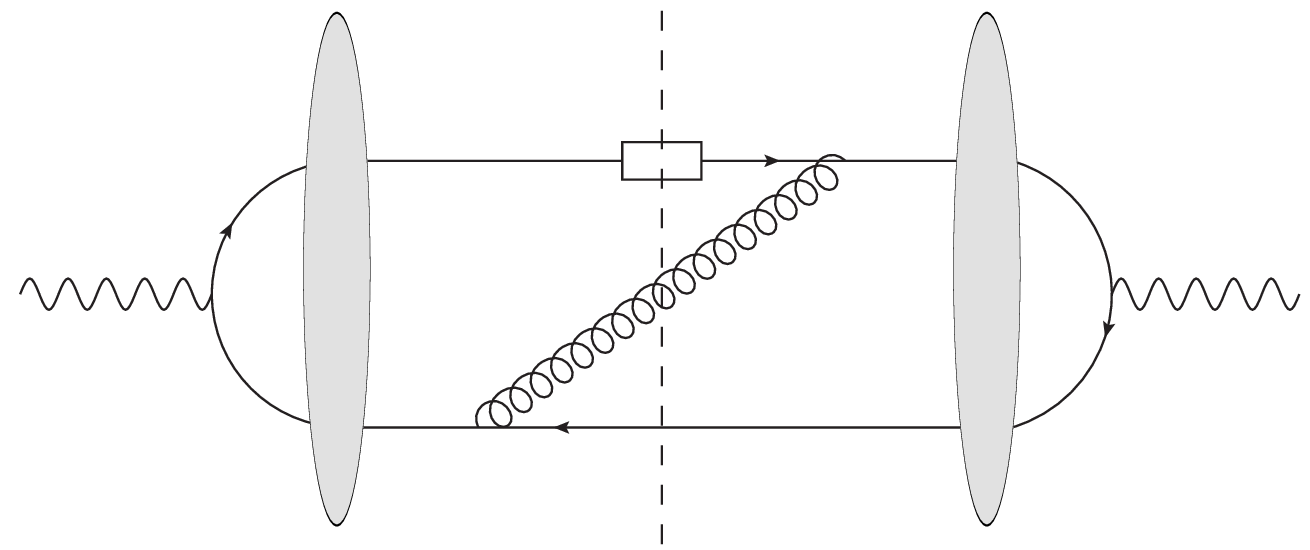}}
\put(322,5){(4)}
\end{picture} 
 \caption{Divergent real contributions to the cross-section, in the quark fragmentation case.}
\label{fig:QuarkInANutshell}
\end{figure}
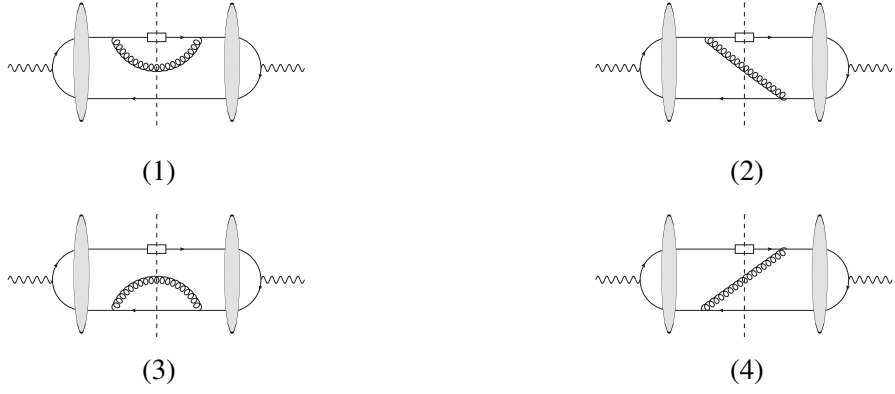
In the case of quark fragmentation, the divergent diagrams associated to real corrections are those shown in Fig.~(\ref{fig:QuarkInANutshell}). Some diagrams posses overlapping soft and collinear singularites that we can disentangle by adding and subtracting the soft limit of each diagrams, i.e,
\begin{equation}
\begin{split}
   & \frac{d \sigma_{ J I}^{q \rightarrow h}}{d x_{h} d^d p_{h \perp} } \bigg |_{\substack{\text{real, singular} \\ \text{dip. $\times$ dip.}}} = \frac{d \sigma_{ J I}^{q \rightarrow h}}{d x_{h} d^d p_{h \perp} } \bigg |_{(1)} + \frac{d \sigma_{ J I}^{q \rightarrow h}}{d x_{h} d^d p_{h \perp} } \bigg |_{(2)} + \frac{d \sigma_{ J I}^{q \rightarrow h}}{d x_{h} d^d p_{h \perp} } \bigg |_{(3)} + \frac{d \sigma_{ J I}^{q \rightarrow h}}{d x_{h} d^d p_{h \perp} } \bigg |_{(4)} \\
        & \equiv \frac{d \sigma_{ J I}^{q \rightarrow h}}{d x_{h} d^d p_{h \perp} } \bigg |_{\text{soft}} + \frac{d \sigma_{ J I}^{q \rightarrow h}}{d x_{h} d^d p_{h \perp} } \bigg |_{\text{coll}(qg)} + \frac{d \sigma_{ J I}^{q \rightarrow h}}{d x_{h} d^d p_{h \perp} } \bigg |_{\text{coll}(\bar{q}g)} + \frac{d \sigma_{ J I}^{q \rightarrow h}}{d x_{h} d^d p_{h \perp} } \bigg |_{\text{real, fin. sub.}} \, .
\end{split}
\label{Eq:Real_Quark}
\end{equation}
The first three terms in the last equality of Eq.~(\ref{Eq:Real_Quark}) are IR-singular. When we combine the singular part of these real corrections with the divergent part of virtual corrections, all soft singularities cancel as expected and we are left with some collinear singularities associated to the final state radiation. These latter are cancelled by part of the IR-singularities of the next-to-leading order renormalized quark $\rightarrow$ hadron FF. In particular, with the part associated to the splitting kernel $P_{qq}$. The cancellation of divergences in the anti-quark channel follows exactly the same path.
\subsection*{Gluon fragmentation}
\begin{figure}[h]
\begin{picture}(430,80)
\put(50,30){\includegraphics[scale=0.18]{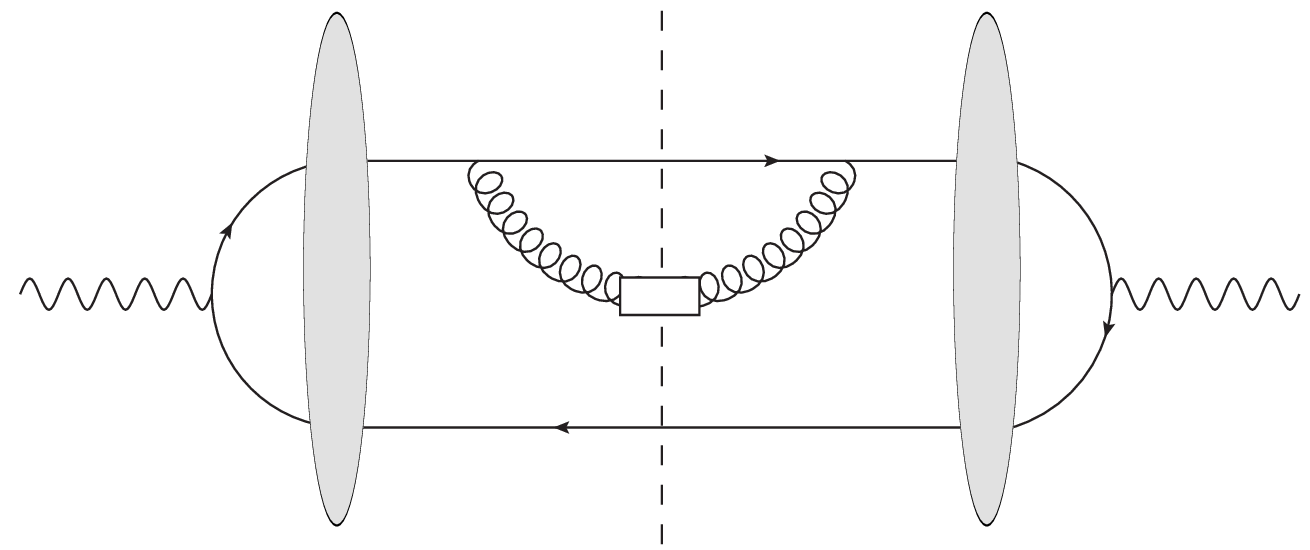}}
\put(102,10){(5)}
\put(270,30){\includegraphics[scale=0.18]{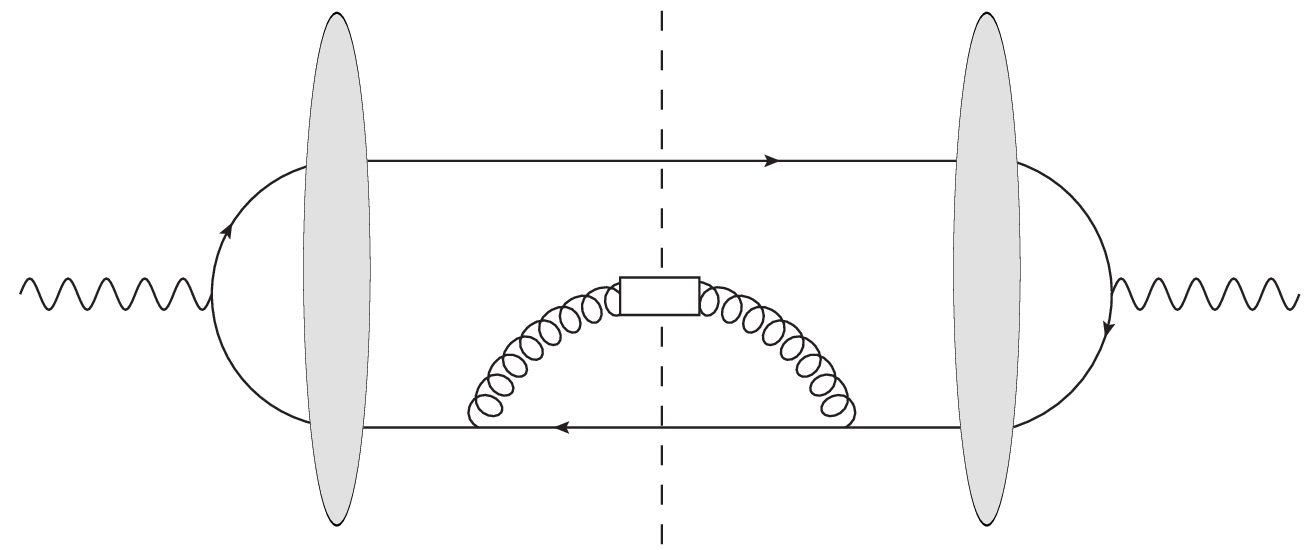}}
\put(322,10){(6)}
\end{picture} 
 \caption{Divergent real contributions to the cross-section, in the gluon fragmentation case.}
\label{fig:GluonInANutshell}
\end{figure}
In the case of gluon fragmentation, the divergent diagrams associated to real corrections are those shown in Fig.~(\ref{fig:GluonInANutshell}) and they are of purely collinear nature. The resulting collinearly divergent contributions to the cross-section, i.e  
\begin{equation}
    \frac{d \sigma_{ J I}^{g \rightarrow h}}{d x_{h} d^d p_{h \perp} } \bigg |_{\text{coll}(qg)} \equiv \frac{d \sigma_{ J I}^{g \rightarrow h}}{d x_{h} d^d p_{h \perp} } \bigg |_{(5)} \; , \hspace{2 cm} \frac{d \sigma_{ J I}^{g \rightarrow h}}{d x_{h} d^d p_{h \perp} } \bigg |_{\text{coll}(\bar{q}g)} \equiv \frac{d \sigma_{ J I}^{g \rightarrow h}}{d x_{h} d^d p_{h \perp} } \bigg |_{(6)} \; .
\end{equation}
are cancelled by the IR-singularities of the part of the NLO renormalized quark and anti-quark FFs associated to the splitting kernel $P_{gq}$. The final finite result of the cross-sections can be found in Refs.~\cite{Fucilla:2022wcg,Fucilla:2023mkl}.  

\section{Acknowledgments}

This  project  has  received  funding  from  the  European  Union’s  Horizon  2020  research  and  innovation program under grant agreement STRONG–2020 (WP 13 "NA-Small-x").
The work by M.~F. is supported by
Agence Nationale de la Recherche under the contract ANR-17-CE31-0019.
The  work of  L.~S. is  supported  by  the  grant  2019/33/B/ST2/02588  of  the  National  Science Center  in  Poland. L.~S. thanks the P2IO Laboratory
of Excellence (Programme Investissements d'Avenir ANR-10-LABEX-0038) and the P2I - Graduate School of Physics of Paris-Saclay University for support.
This work was also partly supported by the French CNRS via the GDR QCD.

\end{document}